\documentstyle[11pt]{article}

\begin{document}
\begin{center}

\renewcommand{\thefootnote}{\fnsymbol{footnote}}

{\LARGE 
The Non-perturbative Effect on  $R=\sigma_L /\sigma_T$\\
\vspace{7mm}

from QCD Vacuum\footnote{This work is supported by National Natural
Science Foundation of China under Grants 19775051 and 
19605006 and 
Natural Science Foundation of Jiansu Province, China.}}
\vfill

{\large Jian-Jun Yang$^{1,3}$, Bo-Qiang Ma$^{2,3}$, Guang-Lie Li$^{2,3}$} 

\vfill
{\small 
$^1$Department of Physics, Nanjing Normal University, Nanjing 210024, China

$^2$CCAST(World Laboratory), P.O.Box 8730, Beijing 100080, China

$^3$Institute of High Energy Physics, Academia Sinica, P.O.Box 918(4), 
Beijing 100039, China\footnote{Mailing address. 
Email address: yangjj@bepc3.ihep.ac.cn; mabq@bepc3.ihep.ac.cn}
}

\vfill

{\large \bf Abstract } \\
\end{center}

\vspace{4mm}
We investigate the non-perturbative effects on the ratio
$R=\sigma_L /\sigma_T$ in lepton-nucleon deep inelastic 
scattering  
by taking into account 
the lowest dimensional condensate 
contributions from the QCD vacuum.  By combining 
conventional perturbative QCD corrections 
and the Georgi-Politzer target-mass
effects
with the non-perturbative effects from the QCD vacuum, 
we give a good description 
of the $Q^2$ and $x$ dependences of $R$ in
comparison with the recent experimental data.

\vfill

\break

\section{Introduction}

The process of lepton-nucleon deep inelastic scattering has proven to 
be an effective tool in 
probing the structure of nucleons. The quantity $R=\sigma_L
/\sigma_T$, the ratio of longitudinal ($\sigma_L$) to transverse
($\sigma_T$) virtual photon absorption cross sections,
is one of the sensitive quantities that can yield 
information about the quark-gluon structure of the nucleon,
thus can provide an additional consistency check of various
theory predictions.
Within the naive parton model with spin-1/2 partons, 
$R$ is expected to be small, and to 
decrease rapidly with increasing momentum transfer $Q^2$, 
whereas with spin-0
partons, $R$ should be large and increase with $Q^2$. 
The earlier measurements of $R$ 
indicated the domination of scattering from spin-1/2 constituents
and confirmed the identification of partons with quarks.

In the naive quark-parton model at finite $Q^2$, 
one can take into account the target-mass effect by the
Callan-Gross relation, i.e. 
$2xF_1=F_2$, and easily find that
\begin{equation}
R(x,Q^2)=\frac{4M^2x^2}{Q^2}.
\end{equation}
In the perturbative QCD, $R$ is expected to decrease 
logarithmically with increasing $Q^2$. However, at finite
$Q^2$ the less known non-perturbative effects and higher twist
effects 
{\it et al.} may also have significant 
consequences on $R$.
Therefore accurate measurements of $R$ with $Q^2$ and
$x$ dependence might be vital for revealing the 
non-perturbative effects and higher twist effects
by the difference between the 
data and the predictions with conventional pQCD corrections
and target-mass effects {\it et al.} taken into account.
There have been some precision measurements of the quantity
$R(x,Q^2)$ in recent years \cite{E140,BCDMS,NMC}, 
and it has been observed
that the data are not completely in compatible with the
predictions with pQCD corrections and target-mass effects
included \cite{E140}. 
This indicates the necessary to take into account
the non-perturbative effects and the higher twist effects {\it et al.}.

The purpose of this paper is to investigate the 
possible non-perturbative
effects in $R(x,Q^2)$.  We will first review
the conventional treatment with pQCD corrections and the 
the Georgi-Politzer (GP) target-mass
effects included. Then we will study the non-perturbative effects on
$R(x,Q^2)$
by taking into account   
the lowest dimensional condensate 
contributions from the QCD vacuum.
It will be shown that a good description 
of the $Q^2$ and $x$ dependences of the data $R$
can be obtained 
by combining the 
pQCD corrections and the Georgi-Politzer (GP) target-mass
effects
with the non-perturbative effects from the QCD vacuum.
 
\section{The pQCD corrections and the Georgi-Politzer mass effects}

We now briefly review
the conventional treatment with the pQCD corrections and the 
Georgi-Politzer (GP) target-mass
effects taken into account. 
The quantity $R$ is defined as the ratio $\sigma_L/\sigma_T$ and 
is related
to the structure functions $F_1(x,Q^2)$ and $F_2(x,Q^2)$ by 
\begin{equation}
R(x,Q^2)=\frac{\sigma_L}{\sigma_T}=\frac{F_2}{2xF_1}(1+\frac{4M^2x^2}{Q^2})-1.
\end{equation}
In the framework of pQCD, logarithmic scaling 
violations \cite{QCDV} occur due to quark-gluon interactions. To the 
order $\alpha_s$, hard gluon bremsstrahlung
from quarks, and photon-gluon interaction effects yield contributions to 
leptonproduction.
The QCD structure functions are given by
\begin{equation}
F_2^{QCD}(x,Q^2)=\sum\limits_i e_i^2 x[q_i(x,Q^2)+q_i(x,Q^2)],
\end{equation}
We use the formula of Ref.~\cite{QCDV} 
for the longitudinal structure function
$F_L$ in next to leading order QCD,
\begin{eqnarray}
F_L^{QCD}(x,Q^2)&=&\frac{\alpha_s(Q^2)}{2\pi}x^2
\int_x^1 \frac{du}{u^3}[\frac{8}{3}F_2^{QCD}(u,Q^2) \nonumber \\
&+&4 \sum\limits_i e_i^2 uG(u,Q^2)(1-x/u)],
\end{eqnarray}
\begin{equation}
2xF_1^{QCD}(x,Q^2)=F_2^{QCD}-F_L^{QCD}
\end{equation}
and
\begin{equation}
R^{QCD}(x,Q^2)=\frac{F_L^{QCD}}{2xF_1^{QCD}}
\end{equation}
where
\begin{equation}
\alpha_s(Q^2)=\frac{12 \pi}{(33-n_f) {\rm ln}[Q^2 / \Lambda^2(n_f)]},
\end{equation}
and $n_f$ is the flavor number of quarks.
The values of $R^{QCD}$, which are shown as dotted curves in
\ref{R}, decrease 
with increasing $Q^2$. 
In addition, the kinematic effects due to target mass, which  
dominate at small $Q^2$ and large $x$, should be further considered.
These effects were first calculated in the framework of operator product
expansion and moment analysis by Georgi and Politzer (GP) \cite{GP}. 
The
structure functions including these GP target-mass effects are given by
\begin{eqnarray}
2F_1^{QTM}(x,Q^2)&=&\frac{1}{k}\frac{2xF_1^{QCD}(\xi,Q^2)}{\xi}+
\frac{2M^2}{Q^2}\frac{x^2}{k^2}I_1 \nonumber \\
& & +\frac{4M^4}{Q^4}\frac{x^3}{k^3}I_2,
\end{eqnarray}
\begin{eqnarray}
F_2^{QTM}(x,Q^2) &=& \frac{x^2}{k^3} \frac{F_2^{QCD}(\xi,Q^2)}{\xi^2}+
\frac{6M^2}{Q^2}\frac{x^3}{k^4}I_1 \nonumber \\
&+&\frac{12M^4}{Q^4} \frac{x^4}{k^5} I_2, 
\end{eqnarray}
and
\begin{equation}
R^{QTM}(x,Q^2)=\frac{F_2^{QTM}}{2xF_1^{QTM}}k^2-1,
\label {QTM}
\end{equation}
where
\begin{equation}
k=(1+\frac{4 x^2 M^2}{Q^2})^{1/2},
\end{equation}
\begin{equation}
\xi=\frac{2x}{1+k},
\end{equation}
\begin{equation}
I_1=\int_\xi^1 du \frac{F_2^{QCD}(u,Q^2)}{u^2},
\end{equation}
and
\begin{equation}
I_2=\int_\xi^1 du \int dv \frac{F_2^{QCD}(v,Q^2)}{v^2}.
\end{equation}

The predicted 
results of $R$ including pQCD corrections and GP target-mass effects 
are also shown as dashed curves in \ref{R}. It has been observed
that the 
results are still significantly lower than the data from recent precision
measurement \cite{E140}. 
This indicates the necessity to take into account other contributions
such as the non-perturbative
effects.

\section{The non-perturbative effects from the QCD vacuum}

We now study the $\left\langle\bar{q}q\right\rangle$-corrected 
quark propagator by taking into account 
the lowest dimensional condensate contributions from the QCD vacuum.
Using the obtained non-perturbative quark propagator, 
we describe the non-perturbative effects on the quantity $R$. 

Let us start by writing the free quark propagator 
\begin{equation}
i[S_F(x-y)]_{\alpha\beta}^{ab}
=\left\langle
0\right|Tq_{\alpha}^{a}(x)\bar{q}_{\beta}^{b}(y)\left|0\right\rangle,
\end{equation}
where $a$, $b$ are color indices 
and $\alpha$, $\beta$ are spinor indices. 
The free quark propagator without any modification of 
condensation can be expressed in momentum space as
\begin{equation}
S_F^{-1}(p)={\not \! p}-m_c 
\end{equation}
with the perturbative (current-) quark mass $m_c$ which can be neglected 
in large momentum transfer process. But in medium energy region, the
quark propagator should be modified by taking into account 
non-perturbative effects \cite{QCDSR,Non-P1,Non-P2}. In 
this paper 
we consider the non-perturbative effects from 
the Feynman diagrams as shown in \ref{QP} 
to the quark propagator.
To  derive 
the effects of dimension-3 quark condensate contribution to the quark  
propagator, we use the non-perturbative vacuum expectation value (VEV) 
of two quark fields 
\begin{equation} 
\left\langle
0\right|\bar{q}_\alpha^{a}(x)q_\beta^{b}(y)\left|0\right\rangle _{NP}
=\frac{1}{12}\delta_{ab}
 (1+\frac{im{\gamma}_{\mu}(x-y)^{\mu}}{4})_{\alpha\beta}
\left\langle\bar{q}q\right\rangle+\cdots
\;.
\end{equation}
For the dimension-4 gluon condensate contribution to the  
quark propagator, we take the non-perturbative VEV of two gluon fields
\begin{equation}
\begin{array}{clcr}
\left\langle0\right|A_\mu^a(x)A_\nu ^b(y)\left|0\right\rangle_{NP}
=\frac{1}{4}x^{\lambda}y^{\rho}
\left\langle0\right|G_{\lambda\mu}^aG_{\rho\nu}^b\left|0\right\rangle
+\cdot\cdot\cdot  \\
=\frac{1}{4}\frac{1}{96}x^{\lambda}y^{\rho}
(g_{\lambda\rho}g_{\mu\nu}-g_{\lambda\nu}g_{\mu\rho})
\left\langle GG \right\rangle+\cdot\cdot\cdot
\end{array}
\end{equation}
in the fixed point gauge $x^{\mu}A_{\mu}(x)=0$,
where 
\begin{equation}  
  \left\langle GG \right\rangle=
\left\langle
0\right|G_{\lambda\mu}^a(0)G_{\lambda\mu}^a(0)\left|0\right\rangle.
\end{equation}
Under the chain approximation, one can obtain the complete quark propagator 
in momentum space \cite{Non-P2}
\begin{equation}
\begin{array}{clcr}
S_F^{-1}(p)={\not \!
p}[1+\frac{g_s^2\left\langle\bar{q}q\right\rangle(1-\xi)m}
{9p^4}+\frac{g_s^2\left\langle GG \right\rangle m_c^2}{12(p^2-m_c^2)^3}]
\\
-[m_c+\frac{g_s^2\left\langle\bar{q}q\right\rangle(4-\xi)}
{9p^2}+\frac{g_s^2\left\langle GG\right\rangle m_cp^2}{12(p^2-m_c^2)^3}],
\end{array}
\end{equation}
where $m$ in $S_F^{-1}(p)$ arises 
from incorporating the QCD equation of motion 
\begin{equation}
(i{\not \! \! D} -m)\psi=0.
\end{equation}
It is necessary to emphasize that $m$, which includes 
the effects of the condensates 
of non-perturbative QCD, is different from 
the purely perturbative (current-) 
quark mass $m_c$. 
Note also that $S_F^{-1}(p)$ is gauge parameter $\xi$ dependent
due to the internal gluon line appearing in \ref{QP}~(b).
In common sense, the current  
quark mass $m_c$ is small, 
and it can be neglected in large momentum transfer process,
which is equivalent to 
neglecting the gluon condensate term in $S_F^{-1}(p)$.
Therefore $S_F^{-1}(p)$ can be rewritten as 
\begin{equation} 
S_F^{-1}(p)={\not \! p}-M(p)
\end{equation}
with
\begin{equation}
M(p)=\frac{g_s^2\left\langle\bar{q}q\right\rangle}
{9p^2}[(4-\xi)-\frac{(1-\xi){\not \! p}m}{p^2}].
\end{equation}

We require the pole of the $\left\langle\bar{q}q\right\rangle$-corrected 
propagator corresponding   
with $m$ in equation of motion (8), i.e.,  
\begin{equation}
M(p)|_{{\not  p}=m}=\frac{g_s^2\left\langle\bar{q}q\right\rangle}{3m^2}=m.
\end{equation}
From this equation, one can obtain the solution of $m$ 
which is independent of gauge parameter
$\xi$ 
\begin{equation}
m=M(p)|_{{\not  p}=m}
=(\frac{4\pi\alpha_s(Q^2)\left\langle\bar{q}q\right\rangle}3)^{1/3}.
\end{equation}
Thus the $\left\langle\bar{q}q\right\rangle$-corrected 
quark propagator can be written as  
\begin{equation} 
S_F^{-1}(p)={\not \! p}-
(\frac{4\pi\alpha_s(Q^2)\left\langle\bar{q}q\right\rangle}3)^{1/3}, 
\label{eq:qp}
\end{equation}

We now discuss the effects of the $\left\langle\bar{q}q\right\rangle$ 
condensate in the 
nucleon structure function by means of 
the non-perturbative quark propagator
Eq.~(\ref{eq:qp}).
Consider the inclusive lepton-nucleon scattering
\begin{equation}
l+N \rightarrow l+X
\end{equation}
where the hadronic structure 
is entirely contained in the tensor $W_{\mu\nu}$
\cite{HKS} 
\begin{equation}
\begin{array}{clcr}
W_{\mu\nu}=(2\pi)^3\sum\limits_{X}
\left\langle P\right|J_{\mu}\left|X\right\rangle\left\langle X\right|
J_{\nu}\left|P\right\rangle\delta^4(P_X-P-q) 
\\ 
=(-g_{\mu\nu}+\frac{q_{\mu}q_{\nu}}{q^2})W_1
+\frac1{M^2}(P_{\mu}-\frac{P.q}{q^2}q_{\mu})
(P_{\nu}-\frac{P.q}{q^2}q_{\nu})W_2,
\end{array}
\end{equation}
here $M$ is the mass of the nucleon.
If $W_{\mu\nu}$ is given, one can extract $W_1$ and $W_2$ 
through the following formulas:
\begin{equation} 
W_1=\frac12[c_2-(1-\frac{\nu^2}{q^2})c_1](1-\frac{\nu^2}{q^2})^{-1};
\label{eq:w1}
\end{equation}
\begin{equation}
W_2=\frac12[3c_2-(1-\frac{\nu^2}{q^2})c_1](1-\frac{\nu^2}{q^2})^{-2},
\label{eq:w2}
\end{equation}
with $c_1\equiv W_{{ } \mu}^\mu$ and
$c_2\equiv \frac{P^{\mu} P^{\nu}}{M^2}W_{\mu\nu}$ \cite{HKS}.
All non-perturbative effects are entirely contained in $W_{\mu\nu}$. 
In this paper, we try to study the non-perturbative effects 
in the nucleon structure function from the quarks 
in the QCD physical vacuum.
We suppose that the proton is made up of bound 
partons that appear as ``free" Dirac particles 
but 
with the non-perturbative 
propagator because of being in the QCD vacuum.
With the incoherence assumption,
one parton contribution to $W_{\mu\nu}$ is  
\begin{equation}
\begin{array}{clcr}
w_{\mu\nu}=(2\pi)^3\frac12\sum\limits_{s,s^{\prime}}
\sum\limits_{p^\prime}
{
\langle
\vec{p},s 
|
J_{\mu}
|
\vec{p^{\prime}},s^{\prime}
\rangle}
{
\langle\vec{p^\prime},s^{\prime}
|
J_{\nu}
|
\vec{p},s
\rangle}
\delta^4(p^\prime-p-q)
\\
=e_i^2\int 
\frac{d^3p^{\prime}}{2p^{\prime}_0}\delta^4(p^{\prime}-p-q)
\frac12 {\rm Tr}\,[\gamma_{\mu}({\not \! p}^{\prime}+m)\gamma_{\nu}
({\not \! p}+m)].
\end{array}
\end{equation}
Neglecting the difference among light quark masses, the quark mass $m$ in 
the  above equation is actually  a c-number multiplied 
by  a unit matrix.
According to the trace theorem that 
the trace of an odd number of $\gamma's$ vanishes,
$w_{\mu\nu}$ can  also be equivalently expressed as
\begin{equation} 
 w_{\mu\nu}=e_i^2\int 
\frac{d^3p^{\prime}}{2p^{\prime}_0}\delta^4(p^{\prime}-p-q)
\frac12 {\rm Tr}\, [\gamma_{\mu}({\not \! p}^{\prime}-m)\gamma_{\nu}
({\not \! p}-m)];
\end{equation}
i.e.,
\begin{equation} 
w_{\mu\nu}=e_i^2\int \frac{d^3p^{\prime}}
{2p^{\prime}_0}\delta^4(p^{\prime}-p-q)
\frac12 {\rm Tr}\,
[\gamma_{\mu}{S_F^{-1}(p^{\prime})}\gamma_{\nu}{S_F^{-1}(p)}],
\label {Wuv}
\end{equation}
\noindent
where $iS_F(p)$ is the quark propagator and $e_i$ is the charge of quark
in unit of $e$. 
Generally, one should take the complete non-perturbative quark propagator
including the corrections  due to $\left\langle\bar{q}q\right\rangle$,
$\left\langle GG\right\rangle$ and higher dimensional condensate. 
However, as a simple qualitative
analysis,
we take only the $\left\langle\bar{q}q\right\rangle$-corrected 
quark propagator given 
by Eq.~(\ref{eq:qp}).
For the sake of simplicity,
we adopt the parton picture in which the parton 4-momentum is expressed as 
$p^\mu=yP^\mu$ $(0\leq y\leq 1)$ with the nucleon 4-momentum $P^\mu$; 
i.e., we 
assume that all transverse momenta are negligible
and that no parton moves oppositely to the nucleon.
Using Eqs.~(\ref{eq:w1})  
and (\ref{eq:w2}), we can extract one quark contribution of type $i$
quark 
to $W_2$, and the corresponding
contribution to the nucleon structure function  $F_2^{(i)}=\nu w_2$ is 
\begin{equation}
F_2^{NP(i)}(y)=2Mx^2e_i^2\delta(y-x)r_2^{NP}(x,Q^2),
\label{eq:Fi}
\end{equation}
with
\begin{equation} 
x=\frac{Q^2}{2M\nu},
\end{equation}
\begin{equation}
r_2^{NP}(x,Q^2)=1+\frac{(3-4 \kappa)\nu}{2MQ^2\kappa^2x}(\frac{4\pi\alpha_s
\left\langle\bar{q}q\right\rangle}{3})^{2/3},
\end{equation}
and 
\begin{equation}
\kappa=1+\frac{Q^2}{4M^2x^2}.
\end{equation}

Suppose that the
nucleon 
state contains $f_i(y){\rm d} y $
parton states of the type $i$ in the interval ${\rm d} y$, then 
\begin{equation}
F_{2}^{NP}=\sum\limits_{i}\int_0^1{\rm d}yf_i(y)F_{2}^{NP(i)}.
\end{equation}
We adopt the convention of Ref.~\cite{HKS}
in which a parton state has $2p_0$ partons per unit volume,
while a nucleon state has $P_0/M$ nucleons per unit volume.
Therefore, in one nucleon, the number of
partons of type $i$, in the interval ${\rm d} y$ is $f_i(y)$ multiplied by 
$\frac{2p_0}{(P_0/M)}=2 M y$,
i.e., $q_i(y){\rm d} y=2 M y f_i(y){\rm d} y$, where $q_i(y)$ is
the quark parton distribution with  
constraint of parton flavor number conservation. 
Summing all contributions of all quarks in the nucleon,
we obtain the structure function of the nucleon
\begin{eqnarray}
F_2^{NP}(x,Q^2)&=&\sum\limits_iq_i(x,Q^2)xe_i^2 r_2^{NP}(x,Q^2)\nonumber \\
               &=&F_2(x,Q^2)r_2^{NP}(x,Q^2)
                 =\sum\limits_i\tilde{q}_i(x,Q^2)xe_i^2,
\label{eq:F2}
\end{eqnarray}
where
\begin{equation}
\tilde{q}_i(x,Q^2)=q_i(x,Q^2)r_2^{NP}(x,Q^2),
\end{equation}
which is different from $q_i(x)$ since $q_i(x)$ represents the 
probability distribution of quarks of type $i$ and satisfies 
the parton
flavor number sum rule, but $\tilde{q}_i(x)$ does not.
From Eq.~(\ref{Wuv}), we can also extract $F_1^{NP}$,
\begin{equation}
\begin{array}{clcr}
F_1^{NP}(x,Q^2)=F_1(x,Q^2)r_1^{NP}(x,Q^2),
\end{array}
\label{eq:F1}
\end{equation}
with
\begin{equation}
r_1^{NP}(x,Q^2)=1+
\frac{(1-4\kappa)
}{\kappa Q^2}
(\frac{4\pi\alpha_s(Q^2)\left\langle\bar{q}q\right\rangle}3)^{2/3}.
\end{equation}

The difference between $r_1^{NP}(x,Q^2)$ and $r_2^{NP}(x,Q^2)$, which is 
due to the non-perturbative effects from QCD vacuum, 
causes the violation of the 
Callan-Gross relation between $F_1$ and $F_2$. 
In Eqs.~(\ref{eq:F2}) and (\ref{eq:F1}),
$F_1$ and $F_2$ are in common sense nucleon structure functions
in terms of realistic quark distributions, which are
taken as those including pQCD corrections and GP target-mass effects,
i.e., 
$F_{1,2}=F_{1,2}^{QTM}$.  
Therefore the 
ratio $R$ including the pQCD corrections, the GP target-mass 
effects and the non-perturbative effects from QCD vacuum
can be written as
\begin{equation}
R(x,Q^2)=\frac {F_2^{QTM}(x,Q^2) r_2^{NP}(x,Q^2)}
{2x F^{QTM}_1(x,Q^2) r_1^{NP}(x,Q^2)}k^2-1.
\label{fR}
\end{equation}

The calculated results of the ratio $R$ based on Eq.~(\ref{fR})
are presented in \ref{R}.
In the estimates of $r_1^{NP}(x,Q^2)$ and $r_2^{NP}(x,Q^2)$, 
we take 
the standard phenomenological 
value of the quark condensate 
$\left\langle\bar{q}q\right\rangle=(250{\rm MeV})^3$ 
obtained from QCD sum
rule \cite{QCDSR}.
We find from {\ref{R}~(a) and (b) that 
the $Q^2$-dependence of $R$ is much improved
as compared with the results in which  
only conventional pQCD corrections 
and target-mass effects are taken into account. 
The results shown in
\ref{R}~(c) and (d) indicate that the x-dependence 
of $R$ is also much better than 
that with only pQCD corrections and target-mass effects.
This reflects the importance of taking into account the
non-perturbative effects, besides the pQCD corrections
and the target-mass effects, for a better 
description of the realistic behaviors of $R$. 

In principle the non-perturbative effects from the QCD vacuum
should also exist in other quantities related to quark distributions, 
such as parton sum rules.
The non-perturbative effects on 
the Gottfried sum rule have been studied 
in a separate work \cite{YML} and 
a non-trivial $Q^2$ dependence beyond the
perturbative QCD is predicted. Unlike the case of the quantity
$R$ considered in this paper, the non-perturbative
effects from QCD vacuum are small in that case and 
do not seem to be the dominant 
source for the violation
of the Gottfried sum rule. 

\section{Summary}

We investigated the non-perturbative effects 
on $R$ by taking into account
the lowest dimensional condensate contributions from the QCD
vacuum in the quark propagator.
We found a non-trivial modification of the conventional quark parton
model formula of the nucleon structure functions at finite $Q^2$ 
and provided a better description of the $Q^2$ and $x$ dependences of 
the recent precision data of $R$
by combining  
conventional perturbative QCD corrections 
and target-mass
effects
with the non-perturbative effects from the QCD vacuum.


\break

\noindent                                                                       
{\large \bf Figure Captions}                                                    
\renewcommand{\theenumi}{\ Fig.~\arabic{enumi}}                                 
\begin{enumerate}                                                               
\item 
The value of $R$ at different $Q^2$ versus $x$: (a). $Q^2=2.5 GeV^2$; 
(b). $Q^2=5.0 GeV^2$. 
The value of $R$ at different $x$ versus $Q^2$: (c). $x$=0.35; 
(d). $x$=0.5. 
The dotted curves are the predictions including only pQCD corrections;
the dashed are the results including 
the pQCD corrections and target-mass effects;
and the  
solid curves are the 
results including also the non-perturbative effects from
QCD vacuum, besides conventional 
pQCD corrections and target-mass effects.
The experimental data are 
taken from Ref.~\cite{E140}.  
The calculations are performed by using the GRV parameterization
\cite{GRV} 
of quark-gluon distribution 
functions.
\label{R}
\vskip 0.5cm
\item                                                                           
The non-perturbative quark propagator including:\\ 
(a). The perturbative free quark propagator;\\
(b). Lowest-order correction due to the nonvanishing  
value of  $\left\langle\bar{q}q\right\rangle$;\\
(c). Lowest-order correction due to the nonvanishing  
value of  $\left\langle GG\right\rangle$.
\label{QP}

\end{enumerate}                                                                 
										
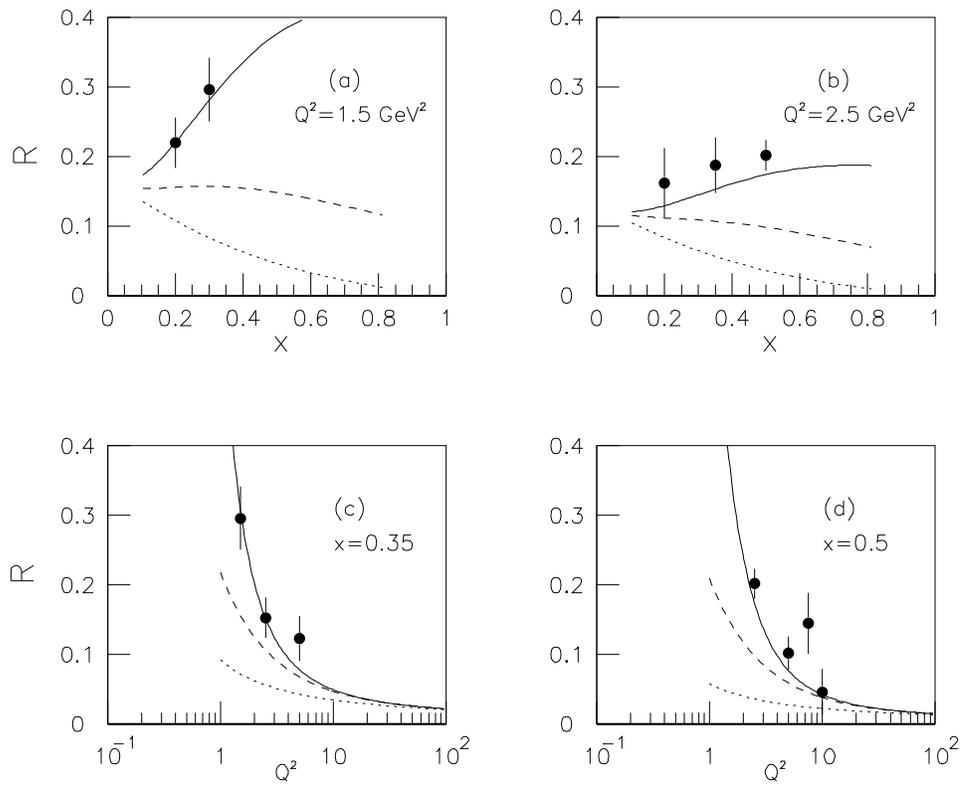
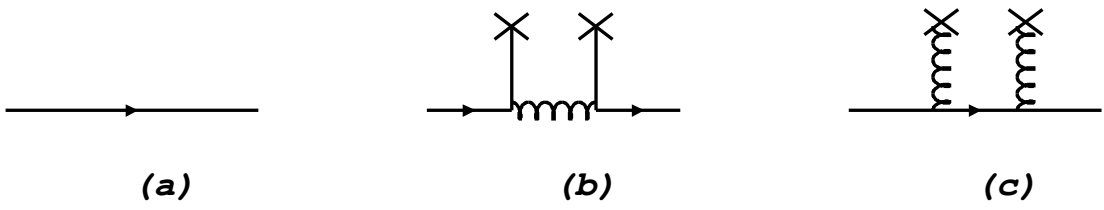
\end{document}